\documentstyle[12pt,epsf]{article}
\begin{document}
\setlength{\baselineskip}{0.33 in}
\catcode`@=11
\long\def\@caption#1[#2]#3{\par\addcontentsline{\csname
  ext@#1\endcsname}{#1}{\protect\numberline{\csname
  the#1\endcsname}{\ignorespaces #2}}\begingroup
    \small
    \@parboxrestore
    \@makecaption{\csname fnum@#1\endcsname}{\ignorespaces #3}\par
  \endgroup}
\catcode`@=12
\newcommand{\newc}{\newcommand}
\newc{\gsim}{\lower.7ex\hbox{$\;\stackrel{\textstyle>}{\sim}\;$}}
\newc{\lsim}{\lower.7ex\hbox{$\;\stackrel{\textstyle<}{\sim}\;$}}
\newenvironment{fnitemize}{\begin{itemize}
	\footnotesize}{\end{itemize}}
\def\NPB#1#2#3{Nucl. Phys. {\bf B#1} #3 (19#2)}
\def\PLB#1#2#3{Phys. Lett. {\bf B#1} #3 (19#2)}
\def\PLBold#1#2#3{Phys. Lett. {\bf#1B} #3 (19#2)}
\def\PRD#1#2#3{Phys. Rev. {\bf D#1} #3 (19#2)}
\def\PRL#1#2#3{Phys. Rev. Lett. {\bf#1} #3 (19#2)}
\def\PRT#1#2#3{Phys. Rep. {\bf#1} #3 (19#2)}
\def\ARAA#1#2#3{Ann. Rev. Astron. Astrophys. {\bf#1} #3 (19#2) }
\def\ARNP#1#2#3{Ann. Rev. Nucl. Part. Sci. {\bf#1} #3 (19#2) }
\def\MODA#1#2#3{Mod. Phys. Lett. {\bf A#1} #3 (19#2) }
\def\ZPC#1#2#3{Zeit. f\"ur Physik {\bf C#1} #3 (19#2) }
\def\APJ#1#2#3{Ap. J. {\bf#1} #3 (19#2) }
\def\bdm{\begin{equation}}
\def\edm{\end{equation}}
\def\beq{\begin{eqnarray}}
\def\eeq{\end{eqnarray}}
\def\bea{\begin{eqnarray}}
\def\eea{\end{eqnarray}}
\def\eac{\Gamma [ \phi_c(x) ]}
\def\ixn{\int d^{4}x_{1} \ldots  d^{4}x_{n}}
\def\opi{\Gamma^{(n)} (x_1,\ldots,x_n )}
\def\vpx{\phi(x)}
\def\vpc{\phi_c(x)}
\def\dpc{{ {\delta \eac} \over {\delta \vpx} }}
\def\vef{V_{\rm{eff}}}
\def\tmx{t_{max}}
\def\lef{\lambda_{\rm{eff}}}
\def\msb{\overline{MS}}
\def\delmu{\partial_{\mu}}
\def\delnu{\partial_{\nu}}
\def\hlf{{1 \over 2}}
\def\tbr{{\overline{t}}_R}
\def\tbl{{\overline{t}}_L}
\def\bbr{{\overline{b}}_R}
\def\bbl{{\overline{b}}_L}
\def\tr{t_R}
\def\tl{t_L}
\def\br{b_R}
\def\bl{b_L}
\def\rxiu{R_{\xi,u}}
\def\rxi{R_{\xi}}
\def\pole{\left[{{-\Delta_{\epsilon}} \over {{\left( 4 \pi \right)}^2}} \right]}
\def\tbr0{{\overline{t}}_{0 R}}
\def\tbl0{{\overline{t}}_{0 L}}
\def\bbr0{{\overline{b}}_{0 R}}
\def\bbl0{{\overline{b}}_{0 L}}
\def\tr0{t_{0 R}}
\def\tl0{t_{0 L}}
\def\br0{b_{0 R}}
\def\bl0{b_{0 L}}

\def\eea{\end{eqnarray}}
\def\sl#1{\setbox0=\hbox{$#1$}           
   \dimen0=\wd0                                 
   \setbox1=\hbox{/} \dimen1=\wd1               
   \ifdim\dimen0>\dimen1                        
      \rlap{\hbox to \dimen0{\hfil/\hfil}}      
      #1                                        
   \else                                        
      \rlap{\hbox to \dimen1{\hfil$#1$\hfil}}   
      /                                         
   \fi}                                         %
%
\vsize 8.7in
\def\singlespace{\baselineskip 11.38 pt}
\def\doublespace{\baselineskip 22.76 pt}
\def\medspace{\baselineskip 17.07 pt}
\font\headings=cmbx10 scaled 1200
\font\title=cmbx10 scaled 1200
\singlespace

\begin{titlepage}
\begin{flushright}
{\large
UP-HEP-9702 \\
hep-ph/9702321\\
February 12, 1997\\
}
\end{flushright}
\vskip 2cm
\begin{center}
{\Large\bf Gauge Dependence of Lower Bounds on the Higgs Mass Derived from Electroweak Vacuum Stability Constraints} 
\vskip 1cm
{\Large
Will Loinaz\footnote{E-mail: {\tt loinaz+@pitt.edu}}\\
R.S. Willey\footnote{E-mail: {\tt willey@vms.cis.pitt.edu}}\\}
\vskip 2pt
{\large\it Department of Physics and Astronomy\\
 University of Pittsburgh, Pittsburgh, PA 15260, USA}\\
\end{center}

\vskip .5cm
\begin{abstract}
We examine the gauge dependence of lower bounds on the 
Higgs mass obtained from the requirement that the electroweak 
vacuum be the global minimum of the effective potential.  
We study a simple model, the
spontaneously-broken Abelian Higgs model coupled to a chiral quark 
doublet in two-parameter $\rxiu$ gauge 
and demonstrate that
the lower bounds on the Higgs mass obtained in this model are dependent on the 
choice of gauge parameters.  We discuss the significance of this result for
calculations in the Standard Model.
\end{abstract}
\end{titlepage}
\setcounter{footnote}{0}
\setcounter{page}{2}
\setcounter{section}{0}
\setcounter{subsection}{0}
\setcounter{subsubsection}{0}
\setcounter{equation}{0}


\input epsf
\section{Introduction}

In the absence of direct observation of the Standard Model Higgs boson,
considerable effort has been expended on the study of
its properties through indirect
means.  There has been a series of papers deriving the lower bound on the
Higgs mass, based on considerations of the stability of the electroweak vacuum 
(see \cite{BigSher} for a comprehensive review of work through 1990).  In the strongest form of the vacuum stability bound, 
it is assumed that the electroweak
vacuum is the absolute minimum of the effective potential ($\vef\left[ \Phi \right]$),
at least up to some `new physics' scale at which the effects of unknown high-scale
physics become significant and the low-energy model is no longer appropriate.
This places restrictions on the running quartic coupling $\lambda$ at the high
scale. After running $\lambda$ down to the electroweak scale, this is converted into a restriction on the Higgs pole mass.

In this paper we investigate the consequences of the gauge dependence of the
effective potential for these calculations.
The gauge dependence of the effective potential was pointed out in
the early 70s \cite{DJ}.  The effective potential is the  sum of one-particle irreducible 
(1-PI) Greens functions at zero external momentum.  In a gauge theory with 
massive
scalars these are offshell quantities, and thus in general gauge-dependent.   
It is that known the value of  $\vef\left[ \Phi \right]$ at any of its extrema, $\Phi_i$, is gauge independent.
However, the locations of these extrema along the $\Phi$ axis, the $\Phi_i$
themselves, are gauge-dependent.  
In practice,
the input condition has been that the renormalization group (RG) improved perturbative $\vef (\Phi)$
should not fall through zero for variable $\Phi$ less than some chosen maximum
(`cutoff')
value $\Phi_{\rm{max}}$. (In some studies, the t'Hooft dimensional regularization
 scale $\kappa$ is identified with
the variable $\Phi$. Then it is $\kappa_{max}$). But the condition
$\vef(\Phi_{\rm{max}})\geq0$
is manifestly gauge dependent (the arbitrarily chosen $\Phi_{\rm{max}}$ is not a
stationary point, so the value of $\vef$ at $\Phi_{\rm{max}}$ depends on the gauge ).
   Alternatively, one could require that the derivative ${{d\vef\left[ \Phi \right]} \over {d \Phi}} $
not go to
zero before $\Phi_{\rm{max}}$, since 
$\left[{{d\vef\left[ \Phi \right]} \over {d \Phi}} \right]_{\Phi_0}=0$  identifies
 $\Phi_0$
as a
stationary point. But the solution $\Phi_0$ is also gauge dependent 
(it is the $vev$ of
a quantum field which is not invariant under the transformations which
have been gauged, e.g. $O(N)$). Thus the proposed condition $\Phi_0$ greater
than or equal to some
arbitrarily fixed $\Phi_{\rm{max}}$ is also gauge dependent.
The relation between the Higgs mass and the scale of new physics,
defined either as a value of $\Phi$ at the extremum of the effective potential
or a value of $\Phi$ at which the effective potential achieves some
fixed value,
relates a physical quantity to an unphysical quantity.  It thus cannot be expected
to be gauge invariant.  

The tree approximation to $\vef$ is gauge invariant.
If one restricts the calculation to the replacement of the
fixed quartic scalar coupling by its RG running version 
in the tree approximation for $\vef$ and uses
a gauge invariant definition of the $\msb$ running masses
and gauge invariant $\beta$ functions, the calculation never
encounters any gauge dependence (i.e. the gauge parameter
$\xi$ never appears).  However, if one attempts to improve the
estimate by including some information from the one-loop 
contributions to $\vef$, the results are essentially infected 
by the gauge dependence described above.  We conclude that the
original estimate cannot be improved.  No model-independent error
estimate is possible in the context of a calculation based on the 
effective potential.  The reader who is convinced by these simple
considerations may skip the lengthy calculations that follow and
skip to section 5 where we discuss some possible
formulations not involving the effective potential which are free of problems
with gauge dependence.

In this paper we demonstrate the gauge-dependence of the lower mass bound
of the Higgs by explicit calculation in a toy model.  Our model consists of a 
spontaneously-broken Abelian Higgs theory coupled to a doublet of chiral fermions
(`top' and `bottom'), one of which
obtains a mass through the spontaneous symmetry breaking.  For certain regions
of the parameter space this model will display the qualitative features of the Standard Model
necessary to study to electroweak vacuum stability.  
In section 2 we outline write out the Lagrangian for the model in a two-
parameter gauge, the $\rxiu$ gauges.  We outline the one-loop $\msb$ 
renormalization of the theory for $\mu^2>0$ and $\mu^2<0$ and give 
an explicit expression for the $vev$ of the Higgs field.  In section 3
we calculate the one-loop effective potential in general $\rxiu$ gauges 
and check that calculating
the location of the extremum of this effective potential gives the same 
$vev$ as the perturbation theory calculation in the broken symmetry phase.
We then do the RG improvement to sum the possibly 
large logarithms introduced by the one-loop contribution to the effective 
potential.
In section 4 we discuss the connection between the RG-improved 
effective potential and the electroweak vacuum instability scale.  We show explicitly
that electroweak vacuum instability scales defined by the location of some feature of
the effective potential along the $\Phi$ axis are gauge-dependent and will pass this 
gauge dependence into a mass bound on the Higgs.
In section 5 we discuss possible alternative methods for deriving lower bounds
on the Higgs mass, and in section 6 we summarize our conclusions. 
\section{The Model}
\subsection{Lagrangian}

The model which we study is the spontaneously-broken
Abelian Higgs model coupled to a chiral fermion doublet.  Over some 
region of its parameter space this model displays the same sort of 
vacuum instability due to heavy fermion loops as arises in the
Standard Model.  We stress that we do {\em not} propose that a numerical lower
bound on the Standard Model Higgs boson be calculated from this model.  It is merely a toy model for illustrating certain issues of principle which 
we wish to present
without the unnecessary complications of a nonabelian gauge theory.  Note in
particular that the fact that a $U\left( 1 \right)$ gauge theory is not 
asymptotically free poses no difficulty as long as vacuum instability scales are
well below the Landau singularities of the couplings.

The Lagrangian is 
\bdm
{\cal{L}}={\cal{L}}_{cl}^{AH}+{\cal{L}}_{cl}^{f}+{\cal{L}}_{gf}+{\cal{L}}_{FP}
\edm
\bea
{\cal{L}}_{cl}^{AH} & = & -{1 \over 4} (\delmu B_{0 \nu} - \delnu B_{0\mu} )^2  + 
            \hlf (\delmu h_{0})^2 + \hlf (\delmu \chi_{0})^2  \nonumber \\	
	& &  + g_{0} \left[ (\delmu h_{0}) \chi_{0} - 
(\delmu \chi_{0}) h_{0} \right] B^{ \mu}_{0}  
	 + \hlf g^{ 2}_{0} B_{0 \mu} B^{\mu}_{0} ( h^{2}_{0} + \chi^{2}_{0}  )  
\nonumber \\
	     & & -	\hlf \mu^{2}_{0}   ( h^{2}_{0} + \chi^{2}_{0}  )    
		- {\lambda_{0}  \over 4} (h^{2}_{0}  +\chi^{2}_{0}  )^2
\eea
\bea
{\cal{L}}_{cl}^{f} & = & i \tbl0 \sl{\partial} \tl0  + i \tbr0 \sl{\partial} \tr0 
+ i \bbl0 \sl{\partial} \bl0 + i \bbr0 \sl{\partial} \br0
  + i g_0 \left[ \tbl0 \sl{B}_0 \bl0 - \bbl0 \sl{B}_0 \tl0 \right]  \nonumber \\
& & -y_{t 0} \left[ \tbl0 \tr0 + \tbr0 \tl0 \right] h_0 - y_{t 0} \left[ \bbl0 \tr0 + \tbr0 \bl0 \right] \chi_0
\eea

As usual, $ \left( h,\chi \right)$ is a doublet of real scalar fields.  $t$ and $b$
are fermion fields.  Their left-handed projections $ \left( t_L,b_L \right)$ form a 
doublet under the local gauge transformation, while their right-handed 
projections transform as a singlet.  Spontaneous symmetry breaking along the 
$h$ direction then gives a tree-level $ \left( \msb \right)$ mass to the Higgs scalar,
the vector boson, and the $t$ fermion.  The Goldstone boson $\chi$ receives
in the broken-symmetry phase no tree-level mass  from ${\cal{L}}_{cl}$, nor does $b$.

To study the gauge-dependence of the mass bound that results from this model,
we choose to work in a two-parameter gauge, the $\rxiu$ gauge. 
The gauge-fixing term is:
\bdm
{\cal{L}}_{gf}= -{1 \over {2 \xi_0}} \left( \partial_{\mu} B_0^{\mu} + \xi_0 u_0 g_0 \chi_0 \right)^2
\edm
and the corresponding Fadeev-Popov ghost term is:
\bdm
{\cal{L}}_{FP}=-{\overline{c}}_0 \sl{\partial} c_0- \xi_0 u_0 g_0^2 {\overline{c}}_0  c_0 h_0
\edm
Special cases of this gauge
include the  Landau gauges $ \left(\xi=0 \right)$,
 't Hooft $R_{\xi}$ gauges \cite{FLS}
($u=v$, the tree-level Higgs $vev$) and Fermi gauges $ \left( u=0 \right)$ .

The gauge-fixing term is chosen anticipating that the spontaneous symmetry
breaking will be in the $h$ direction.  In the $\rxi$ gauges, the gauge-fixing term 
cancels the tree level $\partial B-\chi$ mixing term generated in the 
renormalized classical Lagrangian by the shift of the $h$ field.
  This is especially convenient for perturbative calculations in the
broken-symmetry phase.  The gauge-fixing term explicitly breaks the 
$O\left( 2 \right)$ symmetry of $ {\cal{L}}_{cl}^{AH}$, giving the $\chi$ a tree-level  mass.
The ghost also receives a tree-level mass under SSB.  Except in the 
Fermi and Landau gauges, the ghost is not free.

\subsection{$\msb$ Renormalization}

We renormalize the theory by rescaling the parameters and fields of the 
Lagrangian
by multiplicative renormalization factors $Z$:
\bea
g_0 = Z_g g& \lambda_0  =  Z_{\lambda} \lambda&    y_{t0}  =  Z_{y} y_t  \nonumber\\
\mu^2_0 = Z_{\mu^2} \mu^2 & \xi_0 = Z_{\xi} \xi& u_0=Z_u u \label{eq:couplingZs}
\eea
Among the fields, members of an $O(2)$ doublet receive a common 
renormalization.
\bea
B^{\mu}_0= Z_B^{{1 \over 2}} B^{\mu} &
 \left( h_0, \chi_0 \right)=Z_{\phi}^{{1 \over 2}} \left( h,\chi \right)&
 \left( t_{L0},b_{L0} \right)=Z_L^{{1 \over 2}} \left( t_L,b_L \right) \nonumber \\
t_{R0}= Z_t^{{1 \over 2}} t_R & b_{R0}=Z_b^{{1 \over 2}} b_R
& c_0=Z_c^{{1 \over 2}} c 					\label{eq:fieldZs}
\eea
Writing 
\bdm
Z=1+\left( Z-1 \right)= 1 + \delta Z  \label{eq:Zsep}
\edm
and substituting  (\ref{eq:couplingZs}), (\ref{eq:fieldZs}),
and (\ref{eq:Zsep}) into the Lagrangian generates
the counterterms for the theory.
Various Ward/BRST identities determine that no counterterms are generated
by the renormalization transformations of the gauge-fixing and ghost Lagrangians.

The definition of the counterterms and the physical meaning of the renormalized 
parameters depends on the renormalization scheme.  Here we work in $\msb$.
The counterterms may then be calculated in either the symmetric or broken symmetry phase
of the theory \cite{BWLee}.  

The simplest case, $u=0$ gauges in the symmetric phase $ \left( \mu^2 >0 \right)$
is (aside from the fermions) simply scalar QED (in $O(2)$ rather than $U(1)$
variables).  The $\delta Z$s are calculated
by cancelling the divergences of the two and four point 1-PI functions 
($ {\Gamma}_{\left( n \right)}$).  
The relations between
$Z_g$, $Z_B$ and $Z_{\xi}$ are familiar from QED Ward identities.  A 
particularly
useful feature arises from the relation $Z_{\xi} Z_g^2=1$.  This implies that
$ \xi g^2$ is a renormalization group invariant, although individually the two
factors run.  Thus, once $ \xi g^2$ is fixed to some initial value it remains 
unchanged under RG running.  Since $\xi$ appears only in the combination
$\xi g^2$ in the effective potential, we may simply set the value of $\xi g^2$ 
and never consider the running of $\xi$ alone.

\begin{figure}
\centering
\epsfysize=1.5in  
\hspace*{0in}
\epsffile{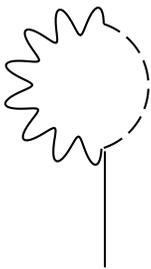} 
\caption{Divergent Higgs One-Point Function for $\mu^2 >0$ in $\rxiu$ Gauge}
\label{mixedonepoint} 
\end{figure}

In the more general $\rxiu$ gauge complications arise as a result of the new pieces of the
gauge-fixing term which explicitly break the global $O\left( 2 \right)$ symmetry
of the ${\cal{L}}_{cl}$.  In particular, the tree-level $\partial B-\chi $ mixing 
induced by this gauge-fixing gives rise to a divergent Higgs one-point function
(see Fig. \ref{mixedonepoint}).
Thus the Higgs field acquires a divergent one-loop $vev$ even for
$\mu^2 >0$.  To perturb about the correct vacuum, 
 we make a
shift of the field
\bdm
h =h^{\prime}+ \delta w
\edm 
and choose $\delta w$ such that 
\bdm
\left\langle 0|h^{\prime}|0 \right\rangle=0.
\edm
Explicitly to one-loop order we find
\bdm
\delta w=-{{\xi g^2 u} \over {\left( 4 \pi \right)^2}} \left[ -\Delta_{\epsilon} +  
\log{{{\mu^2} \over {\kappa^2}}} - 1\right]
\edm
where
\bdm
\Delta_{\epsilon}={2 \over {4-d}} - \gamma_E + \log{4 \pi} 
\edm
is the standard $\msb$ UV subtraction in $d$ dimensions and $\kappa$ is the
renormalization scale.
This shift generates one-loop counterterms which in addition to eliminating the 
one-point  
function of the shifted field $h^{\prime}$ also renders finite the bosonic 
three-point functions (which also arise as a result of $\partial B-\chi$ mixing).  
For $\mu^2 >0$, $\delta w \stackrel{ u \rightarrow 0}{\longrightarrow}0$ and
we return smoothly to the limit of the usual symmetric phase described above.

Perturbation theory for $\mu^2<0$ is most straightforward in the
't Hooft $\rxi$ gauges ($u=v$).  For $\mu^2<0$ the symmetry is 
spontaneously broken already at the tree level.  To perturb about the correct
vacuum we shift the field $h$ by the exact $vev$ $V$.  The exact $vev$ is
determined as a function of the parameters of the theory order-by-order
in perturbation theory by the requirement that the one-point function of the shifted field $\hat{{\Gamma}}_{\left( 1 \right)}$ vanish order-by-order in perturbation
theory.
\bdm
h=\hat{h}+V \Longrightarrow \left\langle 0|\hat{h}|0 \right\rangle=0
\edm
\bdm
V=v+ \delta V
\edm
The gauge-fixing term with $u=v$ cancels the tree-level $\partial B-\chi$
term generated in ${\cal{L}}_{cl}$ by the field shift, leaving $\delta V$
to act as a one-loop counterterm.  The $vev$ $V$ computed to one-loop order
is 
\bea
\lefteqn{V = v+ {v \over {\left( 4 \pi \right)^2}} 
[  \xi g^2 \Delta_{\epsilon} 
-3 \lambda 
\left( \log{{{m_h^2} \over {\kappa^2}}} -1 \right)
- {1 \over 2} {{g^4} \over {\lambda}}
 \left( 3 \log{{{M^2} \over {\kappa^2}}}  -1  \right) } \nonumber \\
& &- {1 \over 2} \xi g^2 
\left( \log{{{\xi M^2} \over {\kappa^2}}} -1 \right) 
+ 2 N_f  {{y_t^4} \over {\lambda}}  
\left( \log{{{m_t^2} \over {\kappa^2}}} -1 \right) ] 
\nonumber\\
& & \label{eq:rxivev}
\eea
In terms of the renormalized parameters of the theory, $\mu^2 \left( <0 \right),
\lambda, g^2$, and $y_t^2$ the masses and tree-level $vev$
 in (\ref{eq:rxivev}) are
\bea
v&=&\sqrt{{{-\mu^2} \over {\lambda}}} \\
m_h^2&=& - 2 \mu^2 \\
M&=&g v \\
m_t&=&y_t v
\eea
Note that $\delta V$ is explicitly gauge-dependent and contains and a UV 
pole $ \left( {\Delta}_{\epsilon} \right) $ \cite{ BWLee,Appelquist,FukadaKugo}.  The subtractions implied by $\delta V$
and the $\delta Z$s eliminate $\hat{\Gamma}_{\left( 1 \right)}$ and render all the other $\hat{\Gamma}_{\left( n \right)}$ finite.

For general $u \neq v$ there is again the complication of uncancelled tree-
level $\partial B-\chi$ mixing, but the procedure is the same.  Again requiring
$\left\langle 0|h^{\prime}|0 \right\rangle  =0$ we calculate the corrections
to the $vev$  
\bea
\lefteqn{    \delta V=-{{\xi g^2 u} \over {\left( 4 \pi \right)^2}} 
[ -\Delta_{\epsilon} + \log{{{\xi g^2 u v} \over {\kappa^2}}} - {3 \over 2} ]   
-{v \over {\left( 4 \pi \right)^2}} [ 3 \lambda 
\left( \log{{{2 \lambda v^2} \over {\kappa^2}}} -1 \right)     } \nonumber \\
& &+ {1 \over 2} {{g^4} \over {\lambda}}
 \left( 3 \log{{{g^2 v^2} \over {\kappa^2}}} - 1 \right) 
- {1 \over 2} \xi g^2 
\left( \log{{{\xi g^2 u v} \over {\kappa^2}}} -2 \right)
- 2 N_f  {{y_t^4} \over {\lambda}}  
\left( \log{{{m_t^2} \over {\kappa^2}}} -1 \right) ]
\eea
Note that for $u=v$ this reproduces the $\rxi$ gauge result (\ref{eq:rxivev}).
Difficulties arise in taking the $u\rightarrow 0$ (Fermi gauge) limit.
IR divergences resulting from the massless Goldstone boson spoil the 
calculation of the one-loop correction to the $vev$.  Fermi gauges are in any case
not especially well-suited to broken-symmetry phase calculations, since the $\partial B-\chi$ 
mixing from ${\cal{L}}_{cl}$ is not cancelled by the gauge-fixing term.  
$\rxiu$ gauges in general also contains $\partial B-\chi$ mixing, but for the gauge
$u=v$ ($\rxi$ gauges)
 the tree-level $\partial B-\chi$ mixing term 
is cancelled.  Thus, Fermi gauge is clearly 
a natural choice for symmetric phase perturbative calculations, as 
$\rxi$ gauges are for broken-symmetry phase calculation.

\section{The Renormalization-Group Improved Effective Potential}

\subsection{One-loop Effective Potential}

The quantity we will refer to as the effective potential, $\vef\left[ \phi \right]$, is the
sum of 1-PI diagrams with any number of external Higgs legs, each carrying
zero external momentum.
\bdm
\vef(\Phi)= -\sum_{n=1}^{\infty} {1 \over {n!}} \Phi^n \Gamma_{(n)}(p_i=0)
\edm
This is in fact not the effective potential, since it is not convex and may be complex.
We sidestep entirely the discussion of convexity, referring the reader to the
literature (e.g. \cite{WeinbergWu}, Appendix of \cite{BigSher}) for the argument that this is indeed the appropriate function to study.   
As mentioned previously, in a theory with massive scalars the
Greens functions at zero external momentum are in general gauge-dependent,
and as a result so is $\vef\left[ \phi \right]$.

The calculation of the one-loop effective potential is most 
simply carried out via the tadpole method  \cite{tadpole}. 
The main complication is that one has to deal with mixed $\partial B-\chi$ 
propagators in any but Landau gauge.
The full one-loop expression is:
\bea
\vef \left[ \Phi \right] & = & {1 \over 2} \mu^2 \Phi^2 +
 {{\lambda} \over 4} \Phi^4 
+ {1 \over 4} {{H^4\left[ \Phi \right]} \over { \left( 4 \pi \right)^2}} 
\left[ \log{{{H^2\left[ \Phi \right]} \over {\kappa^2}}} - {3 \over 2} \right]\nonumber \\
& &+ {3 \over 4} {{B^4\left[ \Phi \right]} \over { \left( 4 \pi \right)^2}} 
\left[ \log{{{B^2\left[ \Phi \right]} \over {\kappa^2}}} - {5 \over 6} \right]
- {2 \over 4} {{G^4\left[ \Phi \right]} \over { \left( 4 \pi \right)^2}} 
\left[ \log{{{G^2\left[ \Phi \right]} \over {\kappa^2}}} - {3 \over 2} \right]\nonumber \\
& &+ {1 \over 4} {{k_+^4\left[ \Phi \right]} \over { \left( 4 \pi \right)^2}} 
\left[ \log{{{k_+^2\left[ \Phi \right]} \over {\kappa^2}}} - {3 \over 2} \right]
+ {1 \over 4} {{k_-^4\left[ \Phi \right]} \over { \left( 4 \pi \right)^2}} 
\left[ \log{{{k_-^2\left[ \Phi \right]} \over {\kappa^2}}} - {3 \over 2} \right] \nonumber \\
& &- N_f {{Y^4\left[ \Phi \right]} \over { \left( 4 \pi \right)^2}} 
\left[ \log{{{Y^2\left[ \Phi \right]} \over {\kappa^2}}} - {3 \over 2} \right]
+\xi g^2 u \Phi \left( \mu^2 + \lambda \Phi^2 \right) \pole  \label{eq:vef}
\eea
where 
\bea
H^2\left[ \Phi \right] &= & \mu^2 + 3 \lambda \Phi^2\\
B^2\left[ \Phi \right]  &= & g^2 \Phi^2\\
G^2\left[ \Phi \right]  &= & \xi g^2 u \Phi\\
k_\pm^2\left[ \Phi \right]   & =& {1 \over 2} \left[ \mu^2 + \lambda \Phi^2 
+ 2 \xi g^2 u \Phi \right]  \pm \nonumber \\
& & {1 \over 2} \sqrt{\left( \mu^2 + \lambda \Phi^2  \right)
\left( \mu^2 + \lambda \Phi^2 + 4 \xi g^2 \Phi \left( u-\Phi \right)  \right)}\\
Y^2\left[ \Phi \right]   & = & y_t^2 \Phi^2
\eea
$H$, $B$, $Y$ and $G$ denote contributions from  Higgs, vector boson, 
heavy fermion and
Fadeev-Popov ghost loops, respectively.  The $k_{\pm}^2$ terms arise
from the $\partial B-\chi$ sector. This result agrees with that
obtained in \cite{AF}, with the exception of a difference in the non-log piece of the B
term \footnote{ We suspect that this is due to a neglected ${{d-4} \over {\epsilon}}=-1 $ contribution in the calculation of the gauge boson
piece.} and the presence of the UV pole.  It is manifestly gauge-dependent,
and the pole term exists for $\xi,u\neq0$.

The expression for the $\vef$ can be checked 
by comparing the expressions for the 
extrema to those obtained by direct perturbative calculation.
The extrema of $\vef$ are found by solving
\bdm
\left[ {{\partial \vef \left[  \Phi \right] } \over {\partial \Phi}} \right]_{\left \langle \Phi \right \rangle} =0
\edm
Writing the extrema of $\vef$ in a loop expansion as
\bdm
\left\langle \Phi \right\rangle=\left\langle \Phi_0 \right\rangle+ 
\left\langle \Phi_1 \right\rangle + \ldots
\edm
this becomes
\bea
0 &{ \stackrel{\rm 1-loop}{=} }&
 \left[ {{\partial \vef^0 \left[  \Phi \right] } \over {\partial \Phi}} \right]_{\left \langle \Phi_0 \right \rangle + \left \langle \Phi_1 \right \rangle} +
 \left[ {{\partial \vef^1 \left[  \Phi \right] } \over {\partial \Phi}} \right]_{\left \langle \Phi_0 \right \rangle} 
\nonumber \\
&=& \left[ {{\partial \vef ^0 \left[  \Phi \right] } \over {\partial \Phi}} \right]_{\left \langle \Phi_0  \right \rangle}  +
\left[ {{\partial^2 \vef ^0 \left[  \Phi \right] } \over {\partial \Phi^2}} \right]_{\left \langle \Phi_0  \right \rangle}  \langle \Phi_1  \rangle+
\left[ {{\partial \vef^1 \left[  \Phi \right] } \over {\partial \Phi}} \right]_{\left \langle \Phi_0 \right \rangle}.
\eea
Solving for the $\left\langle \Phi_0 \right\rangle$ and $\left\langle \Phi_1 \right\rangle$ gives
\bea
\left\langle \Phi_0 \right\rangle &=& \sqrt{{{- \mu^2} \over {\lambda}}} \equiv v\\
\langle \Phi_1 \rangle&=&-\left[ {{\partial^2 \vef ^0 \left[  \Phi \right] } \over {\partial \Phi^2}} \right]^{-1}_{\left \langle \Phi_0  \right \rangle}
\left[ {{\partial \vef^1 \left[  \Phi \right] } \over {\partial \Phi}} \right]_{\left \langle \Phi_0 \right \rangle} = -{1 \over {2 \lambda v^2}} \left[ {{\partial \vef^1 \left[  \Phi \right] } \over {\partial \Phi}} \right]_{\left \langle \Phi_0 \right \rangle}  \nonumber \\
& = & 
-{{\xi g^2 u} \over {\left( 4 \pi \right)^2}} 
[ -\Delta_{\epsilon} + \log{{{\xi g^2 u v} \over {\kappa^2}}} - {3 \over 2} ]   
\nonumber \\
& & -{v \over {\left( 4 \pi \right)^2}} [ 3 \lambda 
\left( \log{{{2 \lambda v^2} \over {\kappa^2}}} -1 \right)     
+ {1 \over 2} {{g^4} \over {\lambda}}
 \left(3 \log{{{g^2 v^2} \over {\kappa^2}}} -1 \right) \nonumber \\
 & & - {1 \over 2} \xi g^2 
\left( \log{{{\xi g^2 u v} \over {\kappa^2}}} -2 \right) 
- 2 N_f  {{y_t^4} \over {\lambda}}  
\left( \log{{{m_t^2} \over {\kappa^2}}} -1 \right) ] 
\eea
which agrees with the perturbative result.
We can also see that the value of $\vef$ at the extremum is 
gauge-invariant:
\bea
\vef \left[ \left\langle \Phi \right\rangle \right] &{ \stackrel{\rm 1-loop}{=} }&
\vef^0 \left[ \left\langle \Phi_0 \right\rangle +  \left\langle \Phi_1 \right\rangle\right] + 
\vef^1 \left[ \left\langle \Phi_0 \right\rangle \right] \nonumber \\
&=& \vef^0 \left[ \left\langle \Phi_0 \right\rangle \right] + 
\left[ {{\partial \vef^0\left[ \Phi \right]} \over {\partial  \Phi}} \right]_{\langle\Phi_0\rangle} 
\left\langle \Phi_1 \right\rangle +
\vef^1 \left[ \left\langle \Phi_0 \right\rangle \right] \nonumber \\
&=& \vef^0 \left[ \left\langle \Phi_0 \right\rangle \right] + 
\vef^1 \left[ \left\langle \Phi_0 \right\rangle \right] \nonumber \\
&=& - {1 \over 4} \lambda v^4 + {1 \over 4} {{m_h^4} \over { \left( 4 \pi \right)^2}} 
\left[ \log{{{m_h^2} \over {\kappa^2}}} - {3 \over 2} \right]\nonumber \\
& &+ {3 \over 4} {{m_B^4} \over { \left( 4 \pi \right)^2}} 
\left[ \log{{{m_B^2} \over {\kappa^2}}} - {5 \over 6} \right]
- N_f {{m_t^4} \over { \left( 4 \pi \right)^2}} 
\left[ \log{{{m_t^2} \over {\kappa^2}}} - {3 \over 2} \right] \label{eq:vefext}
\eea
Upon substituting $\Phi=\left\langle \Phi_0 \right\rangle $ into 
(\ref{eq:vef}) the UV pole disappears, the Fadeev-Popov ghost term
cancels the $k_{\pm}^2$ terms, and the resulting expression (\ref{eq:vefext}) is gauge-independent, as expected.

Observe that in the large-field limit $\Phi \gg u, \mu$ $\vef$ simplifies to 
\bdm
\vef \left[ \Phi \right]  \approx  {1 \over 4 } \Phi^4 \lef\left[ \Phi \right] \label{eq:largefield}
\edm
where
\bea
\lef\left[ \Phi \right] & \equiv & \lambda + \Delta\lambda\left[ \Phi \right] \label{eq:leffbreakup}\\
\Delta\lambda &=&
{{9 \lambda^2} \over { \left( 4 \pi \right)^2}} 
\left[ \log{{{3 \lambda \Phi^2} \over {\kappa^2}}} - {3 \over 2} \right] 
+ {{3 g^4} \over { \left( 4 \pi \right)^2}} 
\left[ \log{{{g^2 \Phi^2} \over {\kappa^2}}} - {5 \over 6} \right] \nonumber \\
& &- N_f {{4y_t^4} \over { \left( 4 \pi \right)^2}} 
\left[ \log{{{y_t^2 \Phi^2} \over {\kappa^2}}} - {3 \over 2} \right] 
+  {{\tilde{k}_+^4} \over { \left( 4 \pi \right)^2}} 
\left[ \log{{{\tilde{k}_+^2 \Phi^2 } \over {\kappa^2}}} - {3 \over 2} \right] \nonumber \\
& &+  {{\tilde{k}_-^4} \over { \left( 4 \pi \right)^2}} 
\left[ \log{{{\tilde{k}_-^2 \Phi^2} \over {\kappa^2}}} - {3 \over 2} \right] 
\label{eq:lambdaeff}
\eea
and
\bdm
\tilde{k}_{\pm}^2= {1 \over 2} \left\{ \lambda \pm 
\sqrt{\lambda \left( \lambda-4 \xi g^2 \right)} \right\} \label{eq:kpm}
\edm

Since $u$ is a gauge parameter, we are free to choose it as large or small as we wish.
We use this freedom to assume $\Phi \gg u$, neglect $u$ and focus on the 
$\xi$ dependence of the quantities of interest.
Since the $u$-dependence has been dropped, this is just what we would
get from a calculation in Fermi gauge (again also neglecting $\mu^2$).
Since $ k_{\pm}^2$ are complex conjugates (as are $ \tilde{k}_{\pm}^2$), their combined contribution
is real regardless the sign of $\lambda$ or $ \lambda-4 \xi g^2$.  However, the
contribution from the Higgs loop becomes complex for $\lambda <0$ if the 
running $ \lambda$ is used inside the one-loop effective potential, as would be the 
case in \cite{CEQ1,CEQ2,CEQ3} had the Higgs loop term been retained.

Taking the  limit ${{\xi g^2} \over {\lambda}} \ll 1,u \ll \Phi$ 
elucidates the structure of the gauge-dependent pieces.  In this limit,
\bea
G^2\left[ \Phi \right]&\sim &0 \\
k_+^2\left[ \Phi \right]&\sim & \lambda\Phi^2 \left[ 1-{{\xi g^2} \over {\lambda}} \right]
-  {{(\xi g^2)^2} \over {\lambda}} \Phi^2 +{\cal{O}}\left( ({\xi g^2})^3 \right) \\
k_-^2\left[ \Phi \right]& \sim & \xi g^2  \Phi^2 +{{(\xi g^2)^2} \over {\lambda}} \Phi^2 + 
{\cal{O}}\left( ({\xi g^2})^3 \right) 
\eea
and 
 \bea
\lef\left[ \Phi \right] & \equiv & \lambda + \Delta\lambda \\
\Delta\lambda &=&
{{9 \lambda^2} \over { \left( 4 \pi \right)^2}} 
\left[ \log{{{3 \lambda \Phi^2} \over {\kappa^2}}} - {3 \over 2} \right] 
+ {{3 g^4} \over { \left( 4 \pi \right)^2}} 
\left[ \log{{{g^2 \Phi^2} \over {\kappa^2}}} - {5 \over 6} \right] \nonumber \\
& &- N_f {{4 y_t^4} \over { \left( 4 \pi \right)^2}} 
\left[ \log{{{y_t^2 \Phi^2} \over {\kappa^2}}} - {3 \over 2} \right] 
+  {{\lambda^2} \over { \left( 4 \pi \right)^2}} 
\left[ \log{{{\lambda \Phi^2 } \over {\kappa^2}}} - {3 \over 2} \right] \nonumber \\
& &-  {{2 \xi g^2 \lambda } \over { \left( 4 \pi \right)^2}} 
\left[ \log{{{\lambda \Phi^2} \over {\kappa^2}}} - 1 \right]
+  {{\left( \xi g^2 \right)^2} \over { \left( 4 \pi \right)^2}} 
\left[ \log{{{\xi g^2} \over {\lambda}}} + {1 \over 2} \right]  \label{eq:deltalambda}
\eea
The $k^2_+$ term carries the piece associated with the Goldstone boson
in Landau gauge.  The $\xi$-dependent terms are of course absent in Landau gauge.

The disconcerting presence of a UV pole in the effective potential 
 is related to the 
renormalization issues discussed previously.  
The effective potential in eq. (\ref{eq:vef}) is expressed as a function of a field that
has had its $Z_{\phi}$ factor removed only.   However, perturbative calculations
indicated that an additional shift of the field is necessary to make the 1-PI 
functions (and thus also $\vef$) finite.  
A shift in the field by the pole piece of $w$ is sufficient to remove the pole
and leave a $\vef$ which is finite.  

It should be noted that the effective potential with
the pole is perfectly well-defined.  A calculation of the extrema of 
$\vef\left[ \Phi \right]$ 
yields a one-loop $vev$ with a pole, but the values of $\vef$ at these
extrema are finite and gauge-independent.  This is to be expected, since
the value of $\vef$ at extrema corresponds to a physical vacuum energy.
However, it is not convenient to work with a divergent $\vef$ for purposes of
studying vacuum stability, and we will choose to work with the expression
for the effective potential in terms of the shifted field, $\Phi^{\prime}=\Phi - 
\delta w_{\rm pole}$.
We might also choose to shift the field by the finite piece, so that the extremum
of $\vef^{\prime}\left[ {\Phi}^{\prime} \right] $ is at $ {\Phi}^{\prime}=0$.  Such
shifts will be unimportant in the $\Phi^{\prime} \gg \mu,u$ limit, however
and we ignore them.

\subsection{Renormalization Group Improvement}

To study  $\vef\left[ \Phi \right]$ for large $\Phi$, we must use the renormalization
group improved effective potential to sum up large logs of the form 
$\log{{{\Phi} \over {\Phi_i}}}$. 
Using the invariance of $\vef\left[ \Phi \right]$ under changes in $\msb$ 
renormalization scale $\kappa$ and dimensional analysis, 
we obtain an equation for the RG-improved $\vef$ that will be valid at large 
fields.  This has the solution
\bdm
{\vef\left( s \Phi_i,\hat{g}_i,\mu_i,\kappa \right)}=\exp{\left[\int_{0}^{\log{s}}{{{4} \over {\gamma_{\phi}\left( x \right) + 1}}} dx\right ]} \vef\left(\Phi_i, \hat{g}\left( s, \hat{g}_i \right), 
\hat{\xi}\left( s, \hat{g}_i,\hat{\xi}_i \right),\mu \left( s, \mu_i \right),
 \kappa \right)
\edm
where
\bdm
\gamma_{\phi}=-{\kappa \over \phi}{ {d \phi} \over {d \kappa}}=
{1 \over 2} {\kappa \over {Z_{\phi}}} {{d Z_{\phi}} \over {d \kappa}}
\edm
 Here $s={{\Phi} \over {\Phi_i} }$,  $\hat{g}$ represents the set of couplings $y_t, \lambda$, and $g$, and $\hat{\xi}$ represents
the pair of gauge parameters $\xi$ and $u$.  

Note that $\mu^2$ and $u$ 
do not appear in the large-field limit of $\vef$ (see eq. (\ref{eq:largefield}) and (\ref{eq:leffbreakup})).  
We also observe that
the gauge parameter appears in $\vef$ only in the combination $\xi g^2$,
at least at one-loop.   The dependence on the gauge parameters 
then reduces to simply the dependence on $\xi g^2$, which is RG invariant.
Thus we can write the RG-improved effective potential for large $\Phi$ as
\bdm
{\vef\left( s \Phi_i,\hat{g}_i,\xi g^2,\kappa \right)}={1 \over 4} 
\lef\left(\Phi_i, \hat{g}\left( s,\hat{g}_i \right),\xi g^2, \kappa \right)
\left( \Phi_i \zeta\left( s \right) \right)^4
\edm
where 
\bdm
\zeta\left( s \right)=\exp{\left[\int_{0}^{\log{s}}{{1 \over {\gamma_{\phi}\left( x \right)+1}}  dx} \right]}
\edm

It has been shown that the $n$-loop effective potential improved 
by $n+1$ loop RGEs resums the $n^{\rm{th}}$-to-leading logs
\cite{Bando, Kastening}.   
This paper does not focus on the resummation of the logs, and 
we are content to sum the leading logs only.  Indeed, the issues
are perhaps clearest in the region in which relatively small field excursions
are necessary and the RG improvement is less important, as is discussed
in the next section.
It is thus sufficient to consider the one-loop effective 
potential with one-loop $\beta$ and $\gamma$ functions.  
In this approximation it is consistent to neglect the running of the couplings $g, y_{t}$,
and $\lambda$ in $\vef^{1}$ and use the 
one-loop running couplings and fields in $\vef^{0}$. 
In this approximation $\Delta\lambda$ is independent of $s$ and $\vef$ is
\bdm
{\vef\left( s \Phi_i,\hat{g}_i,\xi g^2,\kappa \right)}={1 \over 4} 
\left( \lambda \left( s, \hat{g}_i \right) +
\Delta\lambda\left( \Phi_i, \hat{g}_i, \xi g^2,\kappa \right) \right)
\left( \Phi_i \zeta\left( s \right) \right)^4
\edm
where $\Delta\lambda$ is defined in eq.(\ref{eq:deltalambda}).
We note that if the running
$\lambda(s)$ were used in the $\vef^{1}$ the effective potential becomes 
complex for $\lambda(s) <0$.

The $\msb$ RG equations for the running couplings $g^2\left( s,g_i^2\right)$ and
$y_t^2\left( s , \hat{g}_i^2\right)$ can be solved analytically:
\bdm
g^2\left( s,g_i^2 \right)  = {1 \over {{1 \over {g_i^2}}  -  c_2 \log{s}}}  
\edm
where $\beta_{g^2}= c_2 g^4$ and 
\bdm
y_t^2\left( s, \hat{g}_i^2 \right)  = \left[ \left( {{g^2\left( s \right)} \over  {g_i^2} } \right)^{ {c_4}\over {c_2}} \left({1 \over {y^2_i}} - {{c_3} \over {c_2+c_4}} {1 \over {g_i^2}} \right) + 
{{c_3} \over {c_2+c_4}} {1 \over {g^2\left( s \right)}}\right]^{-1}
\edm
where $\beta_{y_t^2}=c_3 y_t^4 - c_4 y_t^2 g^2.$  $y_{ti}^2$ and $g_i^2$ 
are the initial $\left( s=1 \right)$ values of the couplings. 
For $N_f=1$, $c_2={10 \over 3} {1 \over {\left( 4 \pi \right)^2}}$,
$c_3= {{11} \over {\left( 4 \pi \right)^2}}$, and $c_4= {6 \over {\left( 4 \pi \right)^2}}$.
Since this is a $U(1)$ theory and so not asymptotically free, the gauge coupling
exhibits a Landau pole at $s=\exp{\left[{1 \over {g_i^2 c_2}}\right]}.$
 $y_t^2\left( s,\hat{g}_i^2 \right)$
also exhibits a singularity at
\bdm
s=\exp{\left[  {1 \over {g_i^2 c_2}}\right]} \exp{\left[ -{1 \over {c_2}} 
\left\{ \left( g_i^2 \right)^{- { {c_4}\over {c_2 + c_4}}}  \left( {1 \over {g_i^2}} - 
{{c_2+c_4} \over {c_3} } {1 \over {y_{ti}^2}}\right)^{{{c_2} \over {c_2+c_4}}}\right\}\right]}.
\edm
However, this is not relevant to 
our analysis as long as the singularities are far beyond the scale 
at which the electroweak vacuum becomes unstable, 
and can be arranged without difficulty (for the initial parameters we choose here, 
the Landau pole of $g^2(s)$ is at $s=10^{137}$ and the pole in $y^2_{t}(s)$ is 
at $s=5 \times 10^{13}$).
The RG equation for $\lambda\left( s,\hat{g}_i^2 \right)$ must be solved numerically.

\section{The $\msb$ `New Physics' Scale and Lower Bounds on the Higgs Mass
from Vacuum Stability}

As in the Standard Model, $\beta_{\lambda}$ of our model
contains a term due to fermion
loops which tends to drive $\lambda\left(s \right)$ smaller for 
increasing $s$.  For large $y_t^2$ (i.e. for a heavy fermion),
this term may dominate $\beta_{\lambda}$.
At some critical field value of $s$, $\lambda\left( s \right)$ will become
negative, and if this occurs for large $s$ the effective potential 
will quickly become 
much lower than the electroweak minimum.  If our theory were still 
complete at this scale it would imply that the electroweak minimum is
not global minimum of the theory, contrary to our initial assumption.
If we insist that the electroweak vacuum be absolutely stable, 
 we are led to 
conclude that the theory is incomplete at this scale.
Contributions from 
new physics must be significant at this energy scale and either `rescue' the 
effective potential or ruin the entire approach.  

\begin{figure}
\centering
\epsfysize=3.0in  
\hspace*{0in}
\epsffile{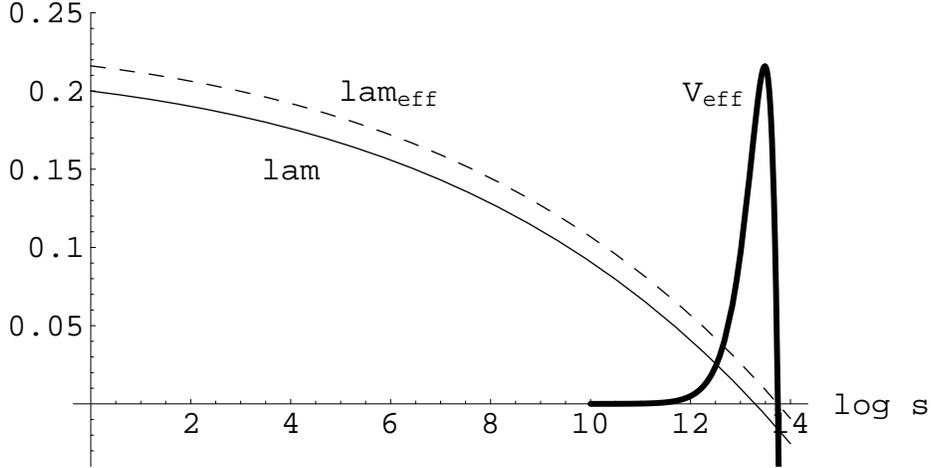} 
\caption{ $\lef\left( s\right)$ (dashed line), $\lambda(s)$
 (solid line), and $\vef\left( s \right)$ (bold line) vs. $ \log{s}$.  \newline
$g^2_i=0.15, y_{ti}^2=0.5, \lambda_i=0.2,\xi g^2=10$.  $\vef$ has
been scaled down to fit the plot.}
\label{wahoo} 
\end{figure}

Previous studies have considered different criteria for instability of 
the electroweak vacuum and corresponding specifications of the instability scale.
Several papers \cite{AI,SherBound,Einhorn} have taken  $\lambda(s_{\rm{max}})=0$ as specifying the vacuum instability scale,
the point at which the RG-improved tree-level effective potential becomes negative.  
Recently Casas, Espinosa, and Quiros \cite{CEQ1,CEQ2,CEQ3} 
have included one-loop corrections to
the Standard Model effective potential (as well as two-loop $\beta$ functions), primarily in an attempt to reduce the renormalization scale ($ \kappa$) 
dependence of the bounds.
They consider the condition $\lef(s_{\rm{max}})=0$, where $\vef(s, \Phi_i)
\approx {1 \over 4} \lef(s) \left( \Phi_i \zeta\left( s \right)\right)^4$, and observe that this
gives a bound on the Higgs mass significantly different from (and 
presumably better than) that from $\lambda(s_{\rm{max}})=0$, at least at low cutoff
scales.  The distinction is illustrated in Fig. \ref{wahoo}.  There $\lef\left( s,\xi g^2,\hat{g}_i \right)$, $\lambda\left( s,\hat{g}_i \right)$, and $\vef$ are
plotted against $\log{s}$.  Observe that $\vef$ falls off sharply, but that
$\lambda$ falls through zero before $\lef$ and $\vef$ do.

The distinction between the conditions $\lambda\left( s \right)=0$ and 
$ \lef\left( s \right)=0$ is
equivalent to the distinction between the conditions $ \vef^0\left( s \right)=0$ and 
$ \vef^0\left( s \right)+\vef^1\left( s \right)=0$.  
$\lambda\left( s\right)$ and $\vef^0$ are gauge independent, 
but the expression for 
$ \lef\left( s \right)$ in eqs. (\ref{eq:lambdaeff}) and (\ref{eq:kpm}) contains 
explicit dependence on the gauge
parameter $\xi$ (the dependence on the gauge parameter $u$ has
dropped out in this approximation).  Thus whether the condition
$\lef\left( s \right)=0$ at some assigned $s_{\rm{max}}$ or the determination
of the instability scale $s_{\rm{max}}$ at which $\lef\left( s \right)$ goes to
zero is used as the vacuum instability criterion, the result will have explicit gauge
$\left( \xi \right)$ dependence if one goes beyond the RG improvement of
the tree-level $\vef$.

To obtain a lower bound on the Higgs pole mass from one of these
stability conditions requires several steps.  As input data one needs
the values of $g_i$ and $y_{ti}$ at some initial low renormalization 
scale $\kappa_i$.  In Standard Model studies these
would come from global fits to electroweak data (with some error estimate)
and introduce no gauge dependence.  Then, using the one-loop approximation
to the $\beta$ functions one has the RG solutions for running $g$ and $y_t$
at any scale (for which they remain perturbatively small).  For convenience
one may take $\kappa_i$ equal to $\Phi_i$, the arbitrary electroweak scale 
at which one specifies the approximate effective potential.
One then integrates the RGE for $\lambda(s)$ starting at $s=1$ ($\Phi=\Phi_i$),
with some initial guess for $\lambda(s=1)$ .  The initial guess is then 
adjusted until the resulting $\lambda\left( s_{\rm{max}} \right)$ at the 
designated cutoff, $\Phi_{\rm{max}}=s_{\rm{max}} \Phi_i$, satisfies the high-scale
boundary condition, $\lambda\left( s_{\rm{max}} \right)=-\Delta\lambda$.
It is at this stage that the gauge dependence enters, through the explicit
$\xi$ dependence of $ \Delta\lambda$.  The resulting critical 
$\lambda\left( s=1 \right)$ becomes the (gauge dependent) lower bound on 
$\lambda\left( \kappa_i \right)$ ($ \lambda\left( \kappa \right)$ is gauge 
independent;  the {\em bound} is gauge dependent).  

Alternately, one could start at the high scale $s_{\rm{max}}$ with the input
$g\left(  s_{\rm{max}}\right)$ and $y_t\left( s_{\rm{max}} \right)$ by running 
the original low scale inputs up to the high scale and by imposing the 
gauge dependent boundary condition on $\lambda\left( s \right)$,
$\lambda\left( s_{\rm{max}} \right)=-\Delta\lambda$.  The RGE for 
$\lambda\left( s \right)$ is then run back down to $s=1$.  The result is the
same--a gauge dependent lower bound on $\lambda\left( \kappa_i \right)$.

There are still some steps to get from
a lower bound on $\lambda\left( \kappa_i \right)$ to a bound on the Higgs 
pole mass, but none of them introduce and new (possibly compensating)
gauge dependence (they do substantially reduce the dependence on the 
arbitrarily-chosen initial renormalization scale $\kappa_i$.)  One requires 
a numerical value for 
$v \left( =\sqrt{{{- \mu^2} \over {\lambda}}}  \right)$ the gauge-independent
tree-level $vev$.  In the Standard Model, $v^2={1 \over {\sqrt{2} G_F }} $
(up to calculated electroweak perturbative corrections).  The other 
ingredient is the relation between the Higgs pole mass $m_h^*$ and
the Higgs $\msb$ mass $m_h^2=2 \lambda v^2$.  This is 
given by the zero of the inverse propagator at $p^2=m_h^{*2}$.  Offshell, the
inverse propagator is explicitly $\xi$ dependent, and it is a nontrivial check
of the calculation that all $\xi$ dependence cancels out onshell. [Note that 
by its definition as the {\em tree level} $vev$ and the renormalization 
conditions chosen, the relation $ v^2= {{m_h^2} \over {2 \lambda}}$
gets no perturbative correction].
\bea
m_h^{*2}&=&m_h^2 + 
 {{\lambda m_h^2} \over {\left( 4 \pi \right)^2}}  \left[ 3 \log{{{m_h^2} \over {\kappa^2}}} +\log{{{M^2} \over {\kappa^2}}} - 12 + 3 \sqrt{3} \pi + 
I\left( {{m_h^2} \over {M^2}} \right) \right]  \nonumber \\
& &+ {{g^2 M^2} \over {\left( 4 \pi \right)^2}}  \left[ 6+6 I\left( {{m_h^2} \over {M^2}} \right) \right] 
 +  {{g^2 m_h^2} \over {\left( 4 \pi \right)^2}}  \left[ -3 \log{{{M^2} \over {\kappa^2}}} + 1 - 2 I\left( {{m_h^2} \over {M^2}} \right) \right] \nonumber \\
& &+  N_f {{y_t^2 m_h^2 } \over {\left( 4 \pi \right)^2}} \left[  2 \log{{{m_t^2} \over {\kappa^2}}}  + 2 I\left( {{m_t^2} \over {m_h^2}} \right) \right]   
- 8 N_f {{y_t^2 m_t^2 } \over {\left( 4 \pi \right)^2}} \left[ I\left( {{m_t^2} \over {m_h^2}} \right)  +1  \right]
\eea
where 
\bdm
I\left( r \right) =\int_{0}^{1}{\log{ \left[1-  r \alpha \left( 1-\alpha \right) \right]} d\alpha}
\edm
In the one-loop corrections to the relation between $m_h^2$ and $m_h^{*2}$
no distinction is made between $m_h^2$ and $m_h^{*2}$.
Thus, the gauge dependence of the lower bound on $\lambda\left( \kappa_i \right)$
obtained from the stability condition propagates into the lower bound on the 
Higgs mass.

\begin{figure}
\centering
\epsfysize=2.5in  
\hspace*{0in}
\epsffile{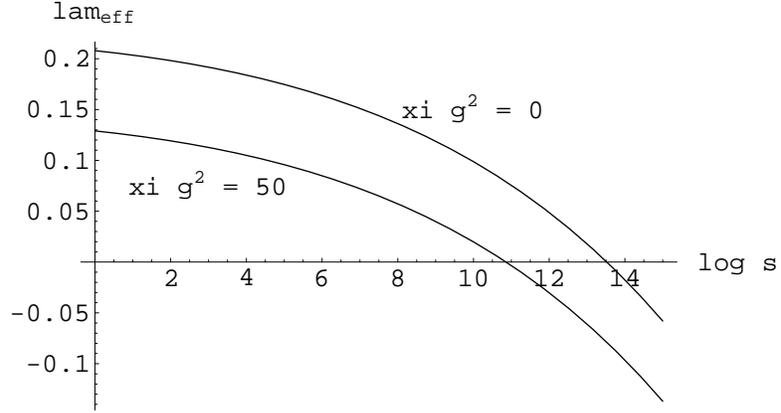} 
\caption{ $\lef\left( s,\xi g^2,\hat{g}_i \right)$ vs. $ \log{s}$ for 
$\xi g^2= 0$ (upper curve) and $\xi g^2=50$ (lower curve).
 $g^2_i=0.15, y_{ti}^2=0.5, \lambda_i=0.2$}
\label{lambdaeffcurves} 
\end{figure}

\begin{figure}
\centering
\epsfysize=2.5in  
\hspace*{0in}
\epsffile{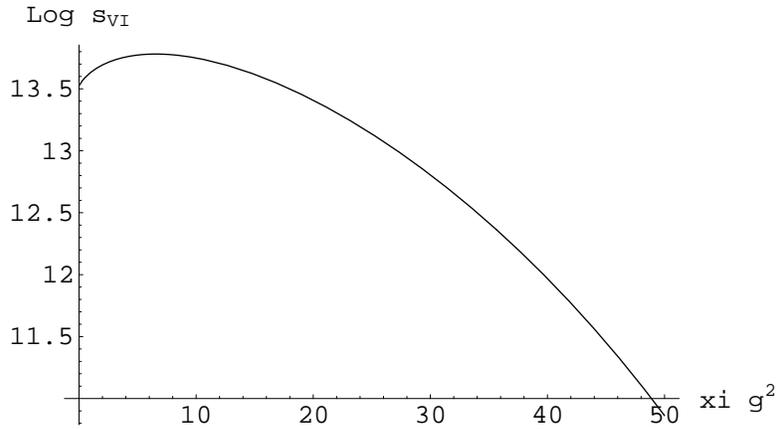} 
\caption{Log of vacuum instability scale vs. $\xi g^2$ for 
 $g^2_i=0.15, y_{ti}^2=0.5, \lambda_i=0.2$.}
\label{leffrootsvsgauge} 
\end{figure}

\begin{figure}
\centering
\epsfysize=2.5in  
\hspace*{0in}
\epsffile{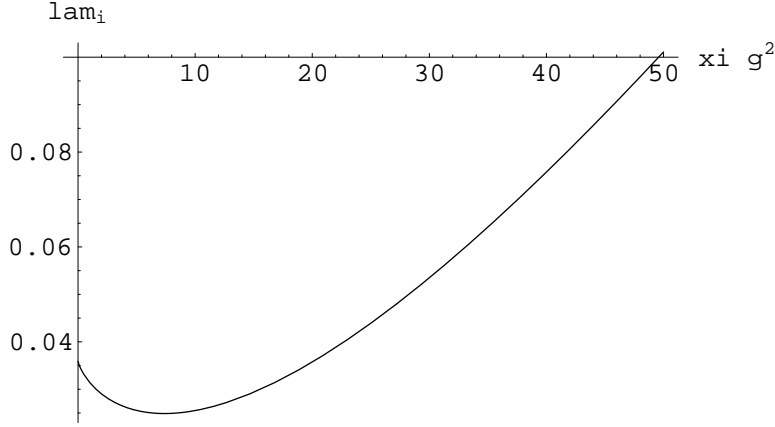} 
\caption{$\lambda_i$ vs. $\xi g^2$ for 
 $g^2_i=0.15, y_{ti}^2=0.5$, assuming a vacuum instability at 
$\log{s_{VI}}=3.7$ and
$\lef(s_{VI})=0$.}
\label{lamivsgauge} 
\end{figure}

To show that the gauge dependence can be made numerically significant, 
in Fig. \ref{lambdaeffcurves} we plot $\lef\left( s \right)$ for several values of
$\xi g^2$.  In Fig. \ref{leffrootsvsgauge} we plot the vacuum instability scale 
corresponding to a particular lower bound on the Higgs pole mass (which
corresponds to some particular $\lambda_i$) as a function of $\xi g^2$.
Following the `top down'  approach in
Fig. \ref{lamivsgauge}, we plot $\lambda_i$ as a function of $\xi g^2$,
assuming that a vacuum instability occurs at $\log{s_{VI}}=3.7.$

We note also that $\vef$ contains additional gauge dependence in the 
scale factor $\zeta\left( s \right)$.  Since this is an overall 
exponential multiplicative
factor, however, it does not change the point at which $\vef=0$.

The explicit effects of the $\xi$-dependent terms on the ''new physics'' scale
are easy to see in the region of parameters space
in which $ \lambda\left( s \right)\approx\lambda_i+
\beta_{\lambda} \log{s}$ is a good approximation.  Solving 
the equation associated with the turnover of $\vef^0$
\bdm
 \lambda\left( s_{\lambda=0} \right)\approx\lambda_i+
\beta_{\lambda} \log{s_{\lambda=0}}=0
\edm
 gives 
\bdm
s_{\lambda=0}=\exp{\left[-{{\lambda_i} \over {\beta_{\lambda}}}\right]}
\edm
Considering instead the condition associated with the one-loop effective potential 
\bdm
\lef\left( s_{\lef=0} \right)=\lambda(s) + \Delta\lambda \approx\lambda_i+
\beta_{\lambda} \log{s_{\lef=0}}+ \Delta\lambda=0
\edm
gives
\bdm
s_{\lef=0}=s_{\lambda=0} \exp{\left[{{-\Delta\lambda} \over {\beta_{\lambda}}}\right]}
\edm
Separating $\Delta\lambda$ into $\xi$-dependent and $\xi$-independent terms
\bdm
\Delta\lambda=\Delta\lambda_{\xi=0}+\delta\lambda_{\xi g^2}
\edm
yields 
\bdm
s_{\lef=0}=s_{\xi=0} \exp{\left[{{-\delta\lambda_{\xi g^2}} \over {\beta_{\lambda}}}\right]}
\edm
That is, the new physics scale can be expressed in terms of the Landau
gauge vacuum instability scale times some gauge-dependent piece which is 
arbitrary.  Thus, we see explicitly that a vacuum instability scale defined in this
manner is necessarily gauge dependent.

More generically, we might propose a vacuum instability scale $\omega$ for 
a gauge choice $\xi$ as the value of the $\Phi$ at which $\vef\left[ \Phi \right]$
achieves some value $c$ ($c$ might be zero, as discussed above, or the
value of  $\vef$ at one of its extrema, or some other numerical value not 
dependent on $\xi$).  
\bdm
\left[ \vef\left[ \Phi,\xi \right] \right]_{\Phi=\omega\left( \xi,c \right)}=c  \label{eq:constraint}
\edm
Any of these possible definitions correspond to different choices of $c$,
but for any $c$ the solution to eq. (\ref{eq:constraint}) $\omega\left( \xi,c \right)$ will be a function of $\xi$.  

\section{Alternative Approaches to the Calculation of Lower Mass Bounds on 
the Higgs}

In view of these problems of gauge dependence of the effective potential 
approach, we might consider other approaches which do not make use
of the effective potential and are manifestly gauge independent.  One such 
approach has already been proposed by one of the authors \cite{Willey}.
That is to just directly solve the (one-loop, coupled) RGEs for the ratio of 
the (gauge independent) running $\msb$ Higgs and top quark masses squared.
Requiring that this remain positive up to some cutoff 't Hooft scale, $\kappa_{\rm{max}}$, again provides a lower bound on the electroweak scale
initial value.  One can include the contributions of the gauge couplings
with no gauge problems, since the $\beta$ functions for the gauge couplings are 
gauge independent.  A problem with this approach is that a negative running 
$\msb$ mass-squared only implies the breakdown of $\msb$ perturbation theory,
not necessarily a disaster of the magnitude of the instability of the vacuum
state in which we live.  Another problem (shared with the minimal, gauge
independent, effective potential approach) is that the connection between the
cutoff 't Hooft scale, $\kappa_{{\rm max}}$ and the masses of the ''new physics'' particles is
not clear.  A simple model in which they are quite different has been give by 
Hung and Sher \cite{HungSher}.

Another possibility is to define a gauge invariant effective potential as the Legendre
transform of a source term coupled to a gauge invariant composite operator
\cite{Banks}.
While this effective potential is gauge invariant, it is not yet determined whether
the problem is computationally tractable.

The problem can also be formulated on a lattice \cite{Lin}.  The Wilson action 
for the gauge fields is gauge invariant and no gauge fixing term is required.
A series of simulations would be run with successively smaller 
values of the input bare $\lambda_0$ and one would look for a nonzero limit
for the output ratio of the Higgs to top masses.  There will again be the 
problem of relating the (lattice) cutoff to a scale of new physics.

\section{Conclusions}

In general it is difficult to extract 
physical information from the effective potential of a gauge theory, and one 
might reasonably be skeptical about the accepting the results of an
effective potential calculation as physical without some concrete demonstration
to that effect. 
One may argue that physical quantities are independent of gauge, and
thus one is free to simply use a convenient gauge.  This is certainly true as 
long as the quantities being calculated are indeed physical quantities in the field theory.
However a {\em bound} on the Higgs pole mass is {\em not} obviously a physical quantity,
and the  fact that this quantity
is gauge-dependent by explicit calculation tells us that in fact it is unphysical,
at least as it arises in the field theory.  
In this case, the expressions for the pole masses of the particles  may indeed be 
expressed in terms of the renormalized 
couplings and mass parameter and have no explicit
gauge dependence order by order in perturbation theory.  The RG equations for
the couplings are also gauge independent.  But the point at which the effective 
potential attains some particular value is not, and using that information to
obtain
numerical values of the Higgs pole mass will inevitably introduce gauge dependence into the number. 
The tree-level $\vef$ does not suffer from gauge dependence, nor will estimates
of the Higgs mass based upon it.  However, 
attempting to incorporate the one-loop $\vef$ to improve these estimate 
inescapably introduces uncontrollable gauge dependence
even as it reduces the renormalization scale dependence.  Hence, no
improvement of the tree-level result will be possible using the effective potential
with current definitions of the vacuum instability scale.
Presumably the issues
raised in this context are also applicable to other problems in which one
assigns physical meaning to features of the 
$\msb$ effective potential of a model containing a 
gauge sector (such as the location of minima).

\section*{Acknowledgments}

We would like to thank Dan Boyanovsky, Tony Duncan, and Marc Sher
for useful conversations.

\section*{Appendix A:  $\msb$ renormalization}

\subsection*{$\msb$ counterterms}
\bea
\delta Z_{\phi}&=&\left[ 2 N_f y_t^2 - 3 g^2  + \xi g^2 \right]\pole  \nonumber \\
\delta Z_{\lambda}&=&\left[ -10 \lambda + 6 g^2 - 3 {{g^4} \over {\lambda}}
- 4 N_f y_t^2 + 4 N_f {{y_t^4} \over {\lambda}} \right]\pole  \nonumber \\
\delta Z_{y}&=&\left[ -y_t^2 \left( {7 \over 4} + N_f \right) + {3 \over 2} g^2 \right]\pole \nonumber \\
\delta Z_{\mu^2}&=&\left[ 3 g^2 - 4 \lambda - 2 N_f y_t^2 \right]\pole \nonumber \\
\delta Z_g&=&- \left[   {{g^2} \over 6} \left( 4 N_f +1  \right) \right]\pole \nonumber\\
\delta Z_{B}&=& \left[   {{g^2} \over 3} \left( 4 N_f +1  \right) \right]\pole \nonumber\\
\delta Z_L&=&\left[ \xi g^2  + {{y_t^2} \over 2} \right]\pole  \nonumber\\
\delta Z_t&=& y_t^2\pole \nonumber \\
\delta Z_c&=&0 \nonumber \\
\delta Z_{\xi}&=& \delta Z_{B}\nonumber\\
\delta Z_u&=&- \delta Z_{\phi}  
\eea

\subsection*{$\beta$ functions}

The relevant one-loop $\msb$ $\beta$ and $\gamma$ functions for the theory are:
\bea
{\beta}_{\lambda} &=&{1 \over {16 \pi^2}} \left( 20 \lambda^2 
- 12 \lambda g^2 + 6 g^4 - 8 N_f y_t^4 + 8 N_f \lambda y_t^2 \right)\\
{\beta}_{g} &=& {1 \over {16 \pi^2}} g^3 \left( {{4 N_f + 1} \over 3 } \right)\\
{\beta}_{y_t} &=& {1 \over {16 \pi^2}} 
\left( \left( {7 \over 4} + N_f  \right) 2 y_t^3 - 3 g^2 y_t^2\right) \\
{\gamma}_{\phi} &=& {1 \over {16 \pi^2}} \left( 2 N_f y_t^2 - 
g^2 \left( 3-\xi \right) \right)
\eea
where $N_f$ is the number of copies of fermion doublet (all assumed to have the 
same couplings).

\vfill\eject

\vfill\eject


\end{document}